\documentclass[aps,prb,12pt,a4paper,preprint,amsmath,amssymb,groupedaddress]{revtex4-1} 

\usepackage{bm}
\usepackage{graphicx}

\usepackage{mathptmx} 

\DeclareSymbolFont{newfont}{OML}{cmm}{m}{it}
\DeclareMathSymbol{\Varrho}{3}{newfont}{37}

\newcommand{\wt}{\widetilde}

\newcommand{\ave}[1]{{\left<#1\right>}}

\newcommand{\abs}[1]{{\left|#1\right|}}

\newcommand{\rmd}{\text{d}}

\newcommand{\taud}{\ensuremath{\tau_\text{d}}}

\newcommand{\tauw}{\ensuremath{\tau_\text{w}}}

\newcommand{\Phiave}{\ensuremath{\ave{\Phi}}}
\newcommand{\Phirms}{\ensuremath{\Phi}_\text{rms}}

\newcommand{\aveA}{\ensuremath{\ave{A}}}

\newcommand{\Eqref}[1]{Eq.~\eqref{#1}}

\newcommand{\Figref}[1]{Fig.~\ref{#1}}

\newcommand{\PPCF}{\textit{Plasma Phys.\ Contr.\ Fusion}}
\newcommand{\PFR}{\textit{Plasma Fusion Res.}}
\newcommand{\PHP}{\textit{Phys.\ Plasmas}}
\newcommand{\PLA}{\textit{Phys.\ Lett.~A}}
\newcommand{\PP}{\textit{Phys.\ Plasmas}}
\newcommand{\PRE}{\textit{Phys.\ Rev.~E}}

\newcommand{\BTSJ}{\textit{Bell Sys.\ Tech. J.}}
\newcommand{\PCPS}{\textit{Proc.\ Cambridge Phil.\ Soc.}}

\newcommand{\PRL}{\textit{Phys.~Rev.\ Lett.}}

\begin{document}

\title{Power law spectra and intermittent fluctuations due to uncorrelated Lorentizan pulses}

\author{O.~E.~Garcia}
\email{odd.erik.garcia@uit.no}
\author{A.~Theodorsen}

\affiliation{Department of Physics and Technology, UiT The Arctic University of Norway, N-9037 Troms{\o}, Norway}

\date{\today}

\begin{abstract}
A stochastic model for intermittent fluctuations due to a super-position of uncorrelated Lorentzian pulses is presented. For constant pulse duration, this is shown to result in an exponential power spectral density for the stationary process. A random distribution of pulse durations modifies the frequency spectrum and several examples are shown to result in power law spectra. The distribution of pulse durations does not influence the characteristic function and thus neither the moments nor the probability density function for the random variable. It is demonstrated that the fluctuations are intrinsically intermittent through a large excess kurtosis moment in the limit of weak pulse overlap. These results allow to estimate the basic properties of fluctuations from measurement data and describe the diversity of frequency spectra reported from measurements in magnetized plasmas.
\end{abstract}

\maketitle



Numerous experimental investigations on fluctuations in magnetized plasmas have demonstrated the general appearance of exponential frequency power spectra and in several cases convincingly attributed this to the presence of Lorentzian pulses in the underlying time series. This includes basic laboratory plasmas,\cite{mm-prl,mm-ppcf,pace-pop,pace-prl} magnetically confined plasmas \cite{mm-ppcf,hornung,mckee} and chaotic and turbulent thermal convection.\cite{mm-pre,garcia-ppcf,garcia-ps} It has been anticipated that Lorentzian pulses and exponential spectra are universal features of pressure-gradient driven turbulence in magnetized plasmas, leading to fluctuation-induced transport.\cite{pace-prl,pace-pop,mm-prl,mm-ppcf,hornung} On the other hand, the frequency spectra of plasma fluctuations are commonly fitted by power laws over limited frequency ranges, obviously motivated by prevailing theories of turbulent motions and self-similar stochastic processes.\cite{pedrosa,sattin,rhodes}

In this contribution, an exponential power spectral density is shown to follow from a super-position of uncorrelated Lorentzian pulses with constant duration. A random distribution of pulse durations significantly influences the frequency spectrum. Several examples of pulse duration distributions resulting in power law spectra are presented, including uniform, Gamma and Rayleigh distributions. The fluctuations are demonstrated to be inherently intermittent, manifested by a large excess kurtosis moment in the limit of weak pulse overlap. Another major conclusion of this work is that the characteristic function and thus the moments and the probability density function for the random variable do not depend on the distribution of pulse duration times. Conversely, the shape of the power spectral density does not depend on the degree of pulse overlap or intermittency.


Consider a stochastic process given by the super-position of a random sequence of $K$ pulses in a time interval of duration $T$,\cite{campbell,rice1,pt,garcia-prl,kg-mse,theodorsen-x,garcia-php,pecseli}
\begin{equation} \label{shotnoise}
\Phi_K(t) = \sum_{k=1}^{K(T)} A_k\varphi\left( \frac{t-t_k}{\tau_k} \right) ,
\end{equation}
where each pulse labeled $k$ is characterized by an amplitude $A_k$, arrival time $t_k$, and duration $\tau_k$, all assumed to be uncorrelated. The pulse arrival times $t_k$ are assumed to be independent and uniformly distributed on the time interval under consideration, that is, their probability density function is given by $1/T$. The pulse duration times $\tau_k$ are assumed to be randomly distributed with probability density $P_\tau(\tau)$, and the average pulse duration time is defined by
\begin{equation}\label{Ptaud}
\taud = \langle \tau \rangle = \int_0^\infty \rmd\tau\,\tau P_\tau(\tau) ,
\end{equation}
where here and in the following, angular brackets denote the average of the argument over all random variables.

The pulse shape $\varphi(\theta)$ is taken to be the same for all events in \Eqref{shotnoise}. This function is normalized such that
\begin{equation} \label{duration}
\int_{-\infty}^{\infty} \rmd\theta\,\abs{\varphi(\theta)} = 1 .
\end{equation}
The integral of an integer power $n$ of the pulse shape will appear frequently in the following, and is defined as
\begin{equation} \label{pulseint}
I_n = \int_{-\infty}^{\infty} \rmd\theta\,\left[ \varphi(\theta) \right]^n .
\end{equation}
It is assumed that $T$ is large compared with the range of values of $t$ for which $\varphi(t/\tau)$ is appreciably different from zero, thus allowing to neglect end effects in a given realization of the process. Furthermore, the normalized auto-correlation function for the pulse shape is defined by
\begin{equation} \label{Rvarphi}
\rho_\varphi(\theta) = \frac{1}{I_2}\int_{-\infty}^{\infty} \rmd\chi\,\varphi(\chi)\varphi(\chi+\theta) ,
\end{equation}
and the Fourier transform of this function is defined by
\begin{equation}
\Varrho_\varphi(\vartheta) = \int_{-\infty}^\infty \rmd\theta\,\rho_\varphi(\theta)\exp(-i\vartheta\theta) .
\end{equation}
In this contribution, the statistical properties of the time series given by \Eqref{shotnoise} will be investigated. In particular, the influence of various duration time distributions in the case of Lorentzian pulses will be explored.


The mean value of $\Phi_K(t)$ is given by averaging over all random variables. Starting with the case of exactly $K$ events in a time interval with duration $T$, this gives \begin{multline} \label{avephiK}
\langle{\Phi_K}\rangle = \int_{-\infty}^\infty \rmd A_1\,P_A(A_1) \int_0^\infty \rmd\tau_1 P_\tau(\tau_1) \int_0^T \frac{\rmd t_1}{T}
\\
\cdots \int_{-\infty}^\infty \rmd A_K\,P_A(A_K) \int_0^\infty \rmd\tau_K\,P_\tau(\tau_K) \int_0^T \frac{\rmd t_K}{T} \sum_{k=1}^K A_k\varphi\left( \frac{t-t_k}{\tau_k} \right) ,
\end{multline}
using that the pulse arrival times are uniformly distributed. Neglecting end effects by taking the integration limits for the arrival times $t_k$ in \Eqref{avephiK} to infinity, the mean value of the signal follows directly,
\begin{equation}
\langle{\Phi_K}\rangle = \taud I_1 \ave{A} \frac{K}{T} .
\end{equation}
Taking into account that the number of pulses $K$ is also a random variable and averaging over this as well gives the mean value for the stationary process,
\begin{equation} \label{phiave}
\ave{\Phi} = \frac{\taud}{\tauw}\,I_1\ave{A} ,
\end{equation}
where $\tauw=T/\langle{K}\rangle$ is the average pulse waiting time. For a non-negative pulse function, $I_1=1$, the mean value of the process is given by the average pulse amplitude and the ratio of the average pulse duration and waiting times. Note that the mean value vanishes both for anti-symmetric pulse shapes, $I_1=0$, and for pulse amplitude distributions with vanishing mean, $\ave{A}=0$. For reasons to become clear presently, the ratio of the average pulse duration and waiting times, $\gamma=\taud/\tauw$, will in the following be refered to as the \emph{intermittency parameter} of the model.\cite{garcia-prl,garcia-php,kg-mse,theodorsen-x}

\begin{figure}
\includegraphics[width=10cm]{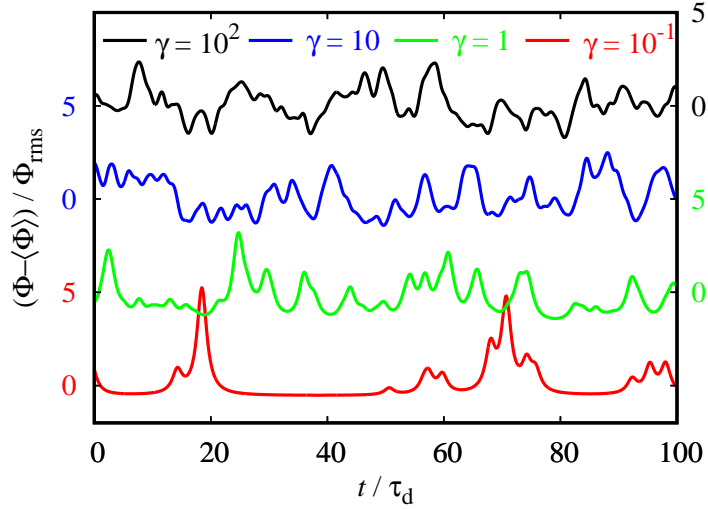}
\caption{Realizations of the stochastic process for Lorentzian pulses with constant duration and exponentially distributed pulse amplitudes. The degree of pulse overlap is determined by the intermittency parameter $\gamma=\taud/\tauw$.}
\label{Lsynthetic}
\end{figure}

The characteristic function for a sum of independent random variables is the product of their individual characteristic functions. Thus, the conditional probability density $P_\Phi(\Phi | K)$ that a sum of $K$ pulse events $\phi_k$ lies in the range between $\Phi$ and $\Phi+\rmd\Phi$ is given by\cite{rice1,garcia-prl,garcia-php,pecseli}
\begin{equation} \label{prbK}
P_\Phi(\Phi | K) = \frac{1}{2\pi}\int_{-\infty}^{\infty} \rmd u\,\exp(-i\Phi u) \prod_{k=1}^K\langle{\exp(i\phi_k u)}\rangle ,
\end{equation}
where the characteristic functions $\langle{\exp(i\phi_ku)}\rangle$ are averaged over the values of $\phi_k=A_k\varphi((t-t_k)/\tau_k)$. For general amplitude and pulse duration distributions,
\begin{equation} \label{charfunc}
\langle{\exp(i\phi_k u)}\rangle = \int_{-\infty}^{\infty} \rmd A_k\,P_A(A_k) \int_0^\infty \rmd\tau_k\,P_\tau(\tau_k) \int_0^T \frac{\rmd t_k}{T} \exp\left[ iuA_k\varphi\left( \frac{t-t_k}{\tau_k} \right)\right] ,
\end{equation}
where $T$ is the duration of the time interval under consideration. Since all the $K$ characteristic functions in \Eqref{prbK} are the same, the conditional probability density is
\begin{equation}
P_\Phi(\Phi | K) = \frac{1}{2\pi}\int_{-\infty}^{\infty} \rmd u\,\exp(-i\Phi u) \langle{\exp(i\phi_k u)}\rangle^K ,
\end{equation}
assuming the number of pulses $K$ in a time interval $T$ to be given. The probability density function for the random variable $\Phi$ is given by
\begin{equation}
P_\Phi(\Phi) = \sum_{K=0}^{\infty} P_\Phi(\Phi|K) P_K(K|T)
= \frac{1}{2\pi}\int_{-\infty}^{\infty} \rmd u\,\exp\left( - i\Phi u + \frac{T}{\tauw}\,\langle\exp(i\phi_k u)\rangle - \frac{T}{\tauw} \right) ,
\end{equation}
where the conditional probability $P_K(K|T)$ that there are exactly $K$ pulse arrivals during any interval of duration $T$ is assumed to be given by the Poisson distribution,
\begin{equation} \label{poisson}
P_K(K|T) = \frac{1}{K!}\left(\frac{T}{\tauw}\right)^K\exp\left(-\frac{T}{\tauw} \right) .
\end{equation}
The stationary probability density function for $\Phi$ is obtained by extending the integration limits for $t_k$ to infinity in \Eqref{charfunc}. This leads to the desired result,
\begin{equation} \label{Pphig}
P_\Phi(\Phi) = \frac{1}{2\pi}\int_{-\infty}^{\infty} \rmd u\,\exp\left( -i\Phi u +
\gamma \int_{-\infty}^{\infty} \rmd A\,P_A(A) \int_{-\infty}^{\infty} \rmd \theta\,\left[ \exp(iuA\varphi(\theta))-1 \right] \right) ,
\end{equation}
which notably is independent of the distribution function for the pulse duration times.
According to this equation, the logarithm of the characteristic function for $P_\Phi$ is
\begin{equation} \label{lncf2}
\gamma \int_{-\infty}^{\infty} \rmd A\,P_A(A) \int_{-\infty}^{\infty} \rmd \theta\,\left[ \exp(iuA\varphi(\theta))-1 \right] = \sum_{n=1}^\infty \gamma I_n\langle{A^n}\rangle \,\frac{(iu)^n}{n!} ,
\end{equation}
where the exponential function on the left hand side has been expanded in a power series. The cumulants $\kappa_n=\gamma I_n\langle{A^n}\rangle$ are the coefficients in the expansion of the logarithm of the characteristic function for $P_\Phi$. From the cumulants, the lowest order moments are readily obtained. A formal power series expansion shows that the characteristic function is related to the raw moments of $\Phi$, defined by $\mu_n'=\ave{\Phi^n}$,
\begin{equation} \label{cfs}
\ave{\exp(i\Phi u)} = 1 + \sum_{n=1}^{\infty} \frac{\ave{i\Phi u}^n}{n!} = 1 + \sum_{n=1}^{\infty} \mu_n'\,\frac{(iu)^n}{n!} .
\end{equation}
It follows that the lowest order centred moments $\mu_n=\ave{(\Phi-\ave{\Phi})^n}$ are related to the cumulants by the relations $\mu_2=\kappa_2$, $\mu_3=\kappa_3$ and $\mu_4=\kappa_4+3\kappa_2^2$. Thus, the variance of the random variable is given by
\begin{equation} \label{phivariance}
\Phirms^2 = \frac{\taud}{\tauw}\,I_2\langle{A^2}\rangle ,
\end{equation}
where $\Phirms$ denotes the standard deviation or root mean square (rms) value of the random variable. Thus, the absolute fluctuation level is large when there is significant overlap of pulses, that is, for long pulse durations and short pulse waiting times. Moreover, the skewness and flatness moments are readily obtained,\cite{garcia-prl,garcia-php}
\begin{subequations} \label{moments}
\begin{align}
S_\Phi & = \frac{\ave{(\Phi-\ave{\Phi})^3}}{\Phirms^3} = \frac{1}{\gamma^{1/2}}\frac{I_3}{I_2^{3/2}}\frac{\langle{A^3}\rangle}{\langle{A^2}\rangle^{3/2}} ,
\\
F_\Phi & = \frac{\ave{(\Phi-\ave{\Phi})^4}}{\Phirms^4} = 3 + \frac{1}{\gamma}\frac{I_4}{I_2^2}\frac{\langle{A^4}\rangle}{\langle{A^2}\rangle^2} .
\end{align}
\end{subequations}
Both these moments increase with decreasing $\gamma$, clearly demonstrating the intrinsic intermittent features of a signal composed of a super-position of pulses. An example of this is clearly seen in \Figref{Lsynthetic}, where time series due to a super-position of Lorentzian pulses with various degree of overlap are presented. Note that for a symmetric amplitude distribution or an anti-symmetric pulse shape, the skewness moment clearly vanishes together with the mean value of the random variable. For large $\gamma$ the skewness and excess flatness moments both vanish, consistent with a normal distribution of the fluctuations which arise in this limit.\cite{rice1,pecseli,garcia-prl,garcia-php} For these reasons, $\gamma$ is referred to as the intermittency parameter of the model. It is emphasized that the distribution of pulse durations does not influence the characteristic function and thus neither the moments nor the probability density function for the random variable.

A general expression for the power spectral density can be derived for this process. Considering first the signal $\Phi_K(t)$ defined by \Eqref{shotnoise}, the auto-correlation function for a given time lag $r$ is given by a double sum comprising $K(K-1)$ terms when $k\neq\ell$ and $K$ terms when $k=\ell$,
\begin{align} \label{PhiKcorr}
\langle{\Phi_K(t)\Phi_K(t+r)}\rangle & = 
\aveA^2\sum_{\substack{k,\ell=1\\ k\neq\ell}}^K \int_0^\infty \rmd\tau_k\,P_\tau(\tau_k) \int_0^T \frac{\rmd t_k}{T}\,\varphi\left( \frac{t-t_k}{\tau_k} \right) \int_0^\infty \rmd\tau_\ell\,P_\tau(\tau_\ell) \int_0^T \frac{\rmd t_\ell}{T}\,\varphi\left( \frac{t-t_\ell+r}{\tau_\ell} \right)
\notag \\
& + \langle{A^2}\rangle \sum_{k=1}^K \int_0^\infty \rmd\tau_k\,P_\tau(\tau_k) \int_0^T\frac{\rmd t_k}{T}\varphi\left( \frac{t-t_k}{\tau_k} \right)\varphi\left( \frac{t-t_k+r}{\tau_k} \right) .
\end{align}
Again neglecting end effects by taking the integration limits for $t_k$ and $t_\ell$ to infinity and averaging over the number of pulses occurring in an interval with duration $T$, it follows that the auto-correlation function for the stationary process is given by
\begin{equation} \label{Rphi}
R_{\Phi}(r) = \langle{\Phi(t)\Phi(t+r)}\rangle = \Phiave^2 + \Phirms^2\,\frac{1}{\taud}\int_0^\infty \rmd\tau\,\tau P_\tau(\tau) \rho_\varphi(r/\tau) .
\end{equation}
From this the power spectral density follows directly by a transformation to the frequency domain,
\begin{equation}
\Omega_\Phi(\omega) = \int_{-\infty}^\infty \rmd r\,R_\Phi(r)\exp(-i\omega r) = 2\pi\Phiave^2\delta(\omega) + \Phirms^2\frac{1}{\taud}\int_0^\infty \rmd\tau\,\tau^2 P_\tau(\tau)\Varrho_\varphi(\omega\tau) ,
\end{equation}
where $\omega$ is the angular frequency and $\delta$ is the delta function. The expression for the frequency spectrum is simplified by considering the centered and scaled random variable $\wt{\Phi}=(\Phi-\Phiave)/\Phirms$,
\begin{equation}
\Omega_{\wt{\Phi}}(\omega) = \frac{1}{\taud}\int_0^\infty \rmd\tau\,\tau^2 P_\tau(\tau) \Varrho_\varphi(\omega\tau) .
\end{equation}
It should be noted that this power spectral density is independent of the amplitude distribution $P_A$ and does not depend on the intermittency parameter $\gamma$, that is, the degree of pulse overlap. Moreover, the above expression is not restricted to a Poisson distribution for the number of pulses $K(T)$. The only assumptions made are that the pulse arrival times have a uniform distribution and that the two lowest order moments of the process are finite.

In the special case of constant pulse duration, $P_\tau(\tau)=\delta(\tau-\taud)$, the expressions for the auto-correlation function and power spectral density become
\begin{subequations} \label{RS_Phi}
\begin{align}
R_{\Phi}(r) & = \Phiave^2 + \Phirms^2\,\rho_\varphi(r/\taud) ,
\\
\Omega_\Phi(\omega) & = 2\pi\Phiave^2\delta(\omega) + \Phirms^2\taud\Varrho_\varphi(\omega\taud) ,
\end{align}
\end{subequations}
that is, they are simply determined by the auto-correlation function for the fixed pulse shape $\varphi(\theta)$. In the following, the frequency spectrum will be calculated for a Lorentzian pulse shape and various distributions of the pulse duration times. The Lorentzian pulse shape is defined by
\begin{equation} \label{lorentzian}
\varphi(\theta) = \frac{1}{\pi}\frac{1}{1+\theta^2} .
\end{equation}
The integral of the $n$-th power of the pulse function is in this case given by
\begin{equation} \label{Inexp}
I_n = \frac{1}{\pi^{n-1/2}}\frac{\Gamma(n-1/2)}{\Gamma(n)} ,
\end{equation}
where $\Gamma$ is the Gamma function. From this it follows that the mean value of the stationary process is given by $\Phiave=\gamma\aveA$, the variance is $\Phirms^2=\gamma\langle{A^2}\rangle/2\pi$, and the normalized auto-correlation function and its transform are
\begin{subequations}
\begin{align} \label{RvarphiL}
\rho_\varphi(\theta) & = \frac{4}{4+\theta^2} ,
\\
\Varrho_\varphi(\vartheta) & = 2\pi\exp(-2\abs{\vartheta}) .
\end{align}
\end{subequations}
In the special case of constant pulse duration it follows that the auto-correlation function is itself a Lorentzian and therefore has algebraic tails,
\begin{equation}\label{RPhiL}
R_\Phi(r) = \Phiave^2 + \Phirms^2\,\frac{4}{4+(r/\taud)^2} ,
\end{equation}
while the power spectral density has an exponential dependence on the frequency,
\begin{equation}\label{SPhiexp}
\Omega_\Phi(\omega) = 2\pi\Phiave^2\delta(\omega) + \Phirms^2 2\pi\taud \exp\left( - 2\taud\abs{\omega} \right) .
\end{equation}
The first term in the above equation results from the mean value of the signal, giving a zero frequency contribution. The second term is the anticipated exponential spectrum for a super-position of uncorrelated Lorentizan pulses.\cite{hornung,mm-prl,pace-pop,pace-prl,mm-ppcf}

\begin{figure}
\includegraphics[width=10cm]{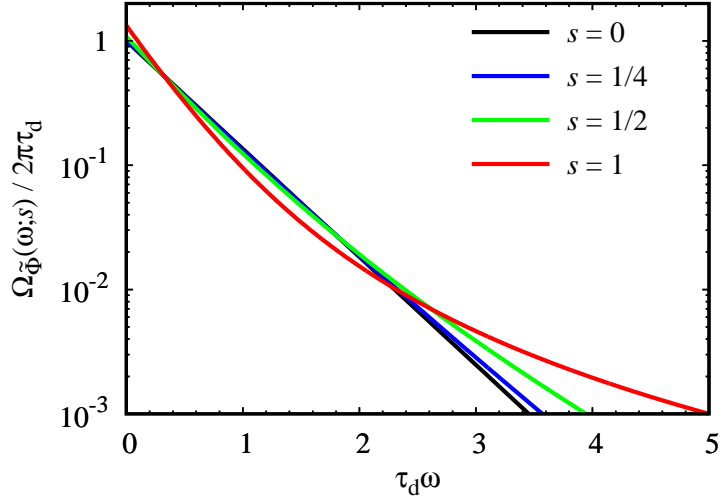}
\caption{Power spectral density for a super-position of Lorentzian pulses with a uniform distribution of duration times with normalized width $s$. The case $s=0$ corresponds to constant pulse duration times, which results in an exponential spectrum.}
\label{SPhi-uniform}
\end{figure}

Any deviation from a constant pulse duration will modify the exponential power spectral density for the random variable. Consider as a first example a uniform distribution of pulse durations, $\taud P_\tau(\tau;s)=1/2s$ for duration times the range $1-s<\tau/\taud<1+s$ and $s$ ranging from zero to unity. In this case the frequency spectrum is given by
\begin{align}
\frac{\Omega_{\wt{\Phi}}(\omega;s)}{2\pi\taud} = \frac{1}{8s\taud^3\abs{\omega}^3} & \left\{ 
[1+2\taud^2\omega^2(1-s)^2 + 2(1-s)\taud\abs{\omega}] \exp[-2(1-s)\taud\abs{\omega}] \right.
\\ \notag
& - \left. [1+2\taud^2\omega^2(1+s)^2 + 2(1+s)\taud\abs{\omega}] \exp[-2(1+s)\taud\abs{\omega}] \right\} .
\end{align}
The power spectral density is presented in a semi-logarithmic plot in \Figref{SPhi-uniform} for various values of the width $s$, clearly showing how the spectrum becomes curved for a broad distribution of pulse duration times.\cite{pace-prl,pace-pop,hornung,mm-ppcf,mm-prl} In the limit $s\rightarrow0$, the pulse duration distribution approaches a delta function, the pulse durations are constant and the exponential spectrum in \Eqref{SPhiexp} is recovered. For the broadest possible distribution of duration times, $s=1$, the above expression simplifies to
\begin{equation}
\frac{\Omega_{\wt{\Phi}}(\omega;1)}{2\pi\taud} = \frac{1-(1+4\taud\abs{\omega}+8\taud^2\omega^2) \exp(-4\taud\abs{\omega})}{8\taud^3\abs{\omega}^3} .
\end{equation}
This spectrum has the asymptotic limits
\begin{subequations}
\begin{gather}
\lim_{\taud\abs{\omega}\rightarrow0} \frac{3}{4}\frac{\Omega_{\wt{\Phi}}(\omega;1)}{2\pi\taud} = 1 ,
\\
\lim_{\taud\abs{\omega}\rightarrow\infty} 8\taud^3\abs{\omega}^3\,\frac{\Omega_{\wt{\Phi}}(\omega;1)}{2\pi\taud} = 1 ,
\end{gather}
\end{subequations}
that is, a flat spectrum for low frequencies and a power law spectrum for high frequencies.

\begin{figure}
\includegraphics[width=10cm]{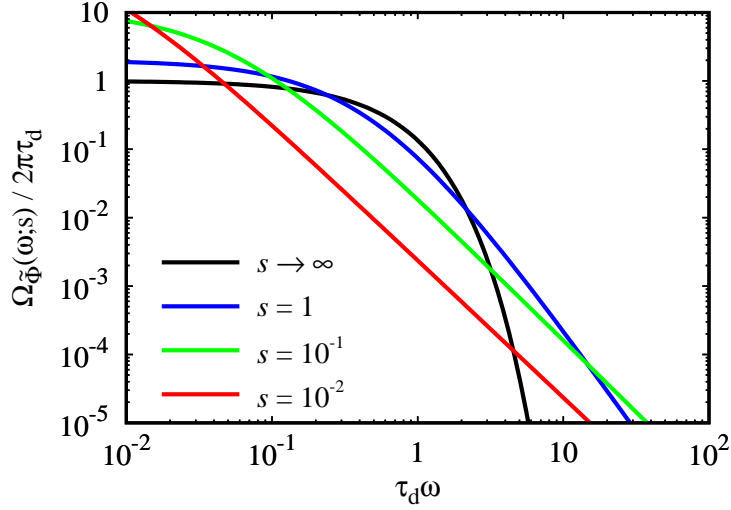}
\caption{Power spectral density for a super-position of Lorentzian pulses with a Gamma distribution of duration times for various shape parameters $s$. The limit $s\rightarrow\infty$ corresponds to constant pulse duration times, which results in an exponential spectrum.}
\label{SPhi-gamma}
\end{figure}

A general probability density function for the duration times is given by the Gamma distribution,
\begin{equation}
\taud P_\tau(\tau;s) = \frac{s^s}{\Gamma(s)}\left( \frac{\tau}{\taud} \right)^{s-1}\exp\left( - \frac{s\tau}{\taud} \right) ,
\end{equation}
with scale parameter $\taud$ and shape parameter $s$. The power spectral density is in this case given by
\begin{equation}
\frac{\Omega_{\wt{\Phi}}(\omega;s)}{2\pi\taud} = \frac{(1+s)s^{1+s}}{(s+2\taud\abs{\omega})^{2+s}} .
\end{equation}
This function is presented in a double-logarithmic plot in \Figref{SPhi-uniform} for various values of the shape parameter $s$. It is of interest to note that for low frequencies, the power spectral density scales with the shape parameter as $(1+s)/s$. Thus, the spectrum becomes increasingly peaked at low frequencies with smaller values of $s$, as is clearly seen in \Figref{SPhi-gamma}. In the limit $s\rightarrow\infty$, the Gamma distribution resembles a narrow normal distribution, corresponding to constant pulse duration. As expected, an exponential power spectral density then results. For $s=1$, the pulse duration times are exponentially distributed, $\taud P_\tau(\tau;1)=\exp(-\tau/\taud)$, and the frequency spectrum is given by
\begin{equation}
\frac{\Omega_{\wt{\Phi}}(\omega;1)}{2\pi\taud} = \frac{2}{(1+2\taud\abs{\omega})^3} ,
\end{equation}
which qualitatively has the same asymptotic limits as spectrum for the broadest possible uniformly distributed pulse duration times discussed above. Finally, in the limit $s\rightarrow0$, yet another power law spectrum results,
\begin{equation}
\lim_{s\rightarrow0} \frac{4\taud^2\omega^2}{s}\,\frac{\Omega_\Phi(\omega;s)}{2\pi\taud} = 1 .
\end{equation}
The examples presented here clearly demonstrate how sensitive the power spectral density is to the distribution of pulse durations. For sufficiently broad duration time distributions, power law spectra are ubiquitous.

\begin{figure}
\includegraphics[width=10cm]{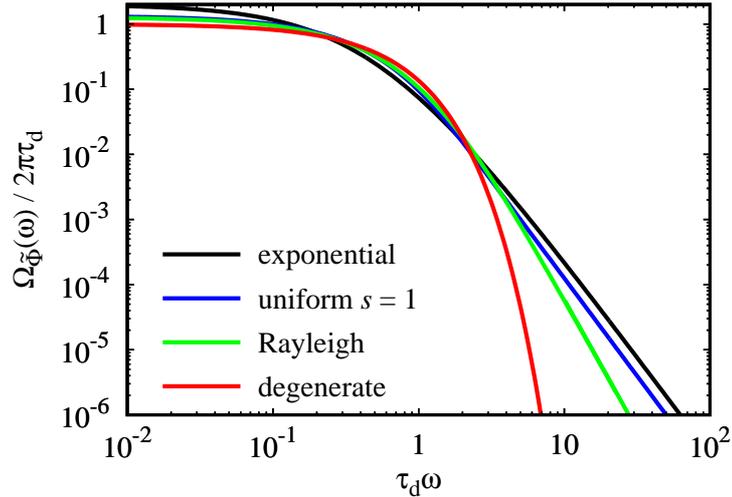}
\caption{Power spectral density for a super-position of Lorentzian pulses with exponential, uniform, Rayleigh and degenerate distribution of pulse duration times.}
\label{SPhi-compare}
\end{figure}

As a final example, consider a Rayleigh distribution of pulse duration times, $\taud P_\tau(\tau)=(\pi\tau/2\taud)\exp(-\pi\tau^2/2\taud^2)$. The power spectral density is then given by
\begin{equation}
\frac{\Omega_{\wt{\Phi}}(\omega)}{2\pi\taud} = \frac{4}{\pi^2} \left[ \pi + 4\taud^2\omega^2 - (3\pi+8\taud^2\omega^2)\taud\abs{\omega}\exp\left( \frac{4\taud^2\omega^2}{\pi} \right) \text{erfc}\left( \frac{2\taud\abs{\omega}}{\pi^{1/2}} \right) \right] .
\end{equation}
This also gives a flat spectrum for low frequencies and a power law spectrum for high frequencies,
\begin{subequations}
\begin{gather}
\lim_{\taud\abs{\omega}\rightarrow0} \frac{\pi}{4} \frac{\Omega_{\wt{\Phi}}(\omega)}{2\pi\taud} = 1 ,
\\
\lim_{\taud\abs{\omega}\rightarrow\infty} \frac{16\taud^4\omega^4}{3\pi} \frac{\Omega_{\wt{\Phi}}(\omega)}{2\pi\taud} = 1 .
\end{gather}
\end{subequations}
In \Figref{SPhi-compare} the power spectral density for the case with constant pulse duration is compared with the cases of exponential, uniform and Rayleigh distributions of pulse durations. All the asymptotic power law limits discussed above are clearly observed.

A reference model for intermittent fluctuations in physical systems has here been extended to include a random distribution of pulse duration times. This is demonstrated to modify the auto-correlation function and power spectral density. In the particularly interesting case of Lorentzian pulse shapes, the power spectrum changes from an exponential function in the case of constant duration times to a power law spectrum for a broad distribution of pulse duration times. By contrast, the characteristic function, and therefore the moments and probability density function, do not depend on the distribution of pulse durations. A robust property of the resulting fluctuations is thus significant skewness and/or excess flatness, in particular in the case of weak overlap of pulse structures. This sheds new light on the statistical properties of fluctuations in physical systems described by such models and allows to estimate the underlying model parameters by using the auto-correlation function or power spectral density and the lowest order moments.

\section*{Acknowledgements}

This work was supported with financial subvention from the Research Council of Norway under grant 240510/F20. The authors acknowledge the generous hospitality of the MIT Plasma Science and Fusion Center where this work was conducted.

\end{document}